# Analytical Investigation of Meson Spectrum via Exact Quantization Rule Approach

Etido P. Inyang[1*], Ephraim P. Inyang[1], Eddy S. William[1], Etebong E. Ibekwe[2] and Ita O. Akpan[1]

[1]Theoretical Physics Group, Department of Physics, University of Calabar, P.M.B 1115, Calabar, Nigeria

[2]Department of Physics, Akwa Ibom State University, Ikot Akpaden, P.M.B 1167, Uyo, Nigeria

Corresponding author email: etidophysics@gmail.com

**Abstract**

We solved the radial Schrödinger equation analytically using the Exact Quantization Rule approach to obtain the energy eigenvalues with the Extended Cornell potential (ECP). The present results are applied for calculating the mass spectra of heavy mesons such as charmonium ($c\bar{c}$) and bottomonium ($b\bar{b}$), and heavy-light mesons such as bottom-charm ($b\bar{c}$) and charm-Strange ($c\bar{s}$) for different quantum states. Two special cases were considered when some of the potential parameters were set to zero, resulting into Coulomb potential, and Cornell potential, respectively. The present potential provides excellent results in comparison with experimental data with a maximum error of $0.0065\ GeV$ and work of other researchers.

## 1. Introduction

The development of the radial Schrödinger equation (SE) in quantum mechanics and its solutions plays a fundamental role in many fields of modern physics. The study of the behavior of quite a lot of physical problems in physics requires the solving of the SE. The solutions can be well-known only if we know the confining potential for a particular physical system [1]. The theory of quantum chromodynamics (QCD) which is described by the meson system is mediated by the strong interactions [2]. The heavy mesons are the constituents of quark and antiquark such as charmonium and bottomonium that are considered as non-relativistic system described by the SE [3]. In recent times, researchers have obtained the solutions of the SE and Klein-Gordon equation (KGE) with the quarkonium interaction potential model such as the Cornell or the Killingbeck potentials [4-13]. The Cornell potential is the sum of the Coulomb plus linear potentials. The

Cornell potential and its extended forms have been solved with SE [14,15] and KGE with different analytical methods [16-18]. The exact solutions of the SE with some potential are solvable for $l = 0$, but insolvable for any arbitrary angular momentum quantum number $l \neq 0$. In this case, several approximate techniques are employed in obtaining the solution. Example of such methods include, asymptotic iteration method(AIM)[6,19] Laplace transformation method [15], super symmetric quantum mechanics method (SUSQM)[20-22], Nikiforov-Uvarov(NU) method [23-34], series expansion method(SEM) [35], analytical exact iterative method(AEIM)[8], WKB approximation method [36] and others[37].

Recently, the mass spectrum of the quarkonium system have been studied by researchers [36,18,35]. For instance, Vega and Flores [38] obtained the solution of the SE with the Cornell potential via the variational method and super symmetric quantum mechanics (SUSYQM). Ciftci and Kisoglu [6] addressed non-relativistic arbitrary $l$-states of quark-antiquark through the Asymptotic Iteration Method (AIM). An analytic solution of the N-dimensional radial SE with the mixture of vector and scalar potentials via the Laplace transformation method (LTM) was studied by [15]. Their results were employed to analyze the different properties of the heavy-light mesons. Also, Al-Jamel and Widyan [9] studied heavy quarkonium mass spectra in a Coulomb field plus quadratic potential by employing the Nikiforov-Uvarov(NU) method. In their work, the spin-averaged mass spectra of heavy quarkonia in a Coulomb plus quadratic potential is analyzed within the non-relativistic SE. In addition Al-Oun et al. [39] examined heavy quarkonia characteristics in the general framework of a non-relativistic potential model consisting of a Coulomb plus quadratic potential. Furthermore, Omugbe et al.[36] solved the SE with Killingbeck potential plus an inversely quadratic potential model via the WKB method. They obtained the energy eigenvalues and the mass spectra of the heavy and heavy-light meson systems. In addition, Inyang et al.[18] obtained the KGE solutions for the Yukawa potential via the NU method. They obtained energy eigenvalues both in relativistic and non-relativistic regime, and the results were then applied to calculate heavy-meson masses of charmonium $c\bar{c}$ and bottomonium $b\bar{b}$. Ibekwe et al.[35] solved the radial SE with an exponential, generalized, harmonic Cornell potential via the series expansion method. They applied the bound state eigenvalues to study the energy spectra for CO, NO, CH and $N_2$ diatomic molecules and the mass spectra of heavy quarkonium systems.

Therefore, in this present work, we aim at studying the SE with the extended Cornell potential via the Exact quantization rule (EQR) to obtain the mass spectra of heavy mesons such as charmonium $(c\bar{c})$, bottomonium $(b\bar{b})$ and the heavy–light mesons such as the charm-Strange $(c\bar{s})$ and bottom-charm $(b\bar{c})$. The extended Cornell potential (ECP) takes the form [12,36].

$$V(r) = \eta_0 r^2 + \eta_1 r - \frac{\eta_2}{r} + \frac{\eta_3}{r^2} \quad (1)$$

where $\eta_0$, $\eta_1$, $\eta_2$ and $\eta_3$ are potential strength parameters. The second term in Eq. (1) is a linear term for confinement feature and the third term is the Coulomb potential that describes the short distance between quarks. While the first and the last terms are quadratic and the inverse quadratic potentials. It is important to note that if we set $\eta_0 = \eta_1 = \eta_3 = 0$, the (ECP) reduces to the Coulomb potential also if we set $\eta_0 = \eta_3 = 0$ the (ECP) reduces to the standard Cornell potential. The paper is organized as follows: in section 2, the brief EQR formalism is presented. Section 3, the analytical solution of the bound states of the SE is solved via the EQR. In section 4, we present the results of the mass spectrum. Finally, in section 5, the paper is concluded.

## 2. Exact Quantization Rule formalism

In this section, we give a brief review of exact quantization rule. The details can be found in [40.41]. It is a well known fact that, in one dimension, the SE is given as:

$$\frac{d^2 \psi(x)}{dx^2} + \frac{2\mu}{\hbar^2}[E_{nl} - V(x)]\psi(x) = 0 \quad (2)$$

Equation (2) can be written in the following form:

$$\phi'(x) + \phi(x)^2 + k(x)^2 = 0 \quad (3)$$

with

$$k(x) = \sqrt{\frac{2\mu}{\hbar^2}[E_{nl} - V(x)]} \quad (4)$$

where $\phi(x) = \psi'(x)/\psi(x)$ is the logarithmic derivative of the wave function, $\mu$ is the reduced mass of the quarkonium particles, $k(x)$ is the momentum, and $V(x)$ is a piecewise continuous real potential function of $x$. The phase angle of the SE is the logarithmic derivative $\phi(x)$. From Eq. (3), as $x$ increases across a node of wave function $\psi(x)$, $\phi(x)$ decreases to $-\infty$, jumps to $+\infty$, and then decreases again. We can generalized EQR to the three – dimensional radial SE with spherically symmetric potential by simply making the replacement $x \to r$ and $V(x) \to V_{eff}(r)$ [40,41].

$$\int_{r_a}^{r_b} k(r)dr = N\pi + \int_{r_a}^{r_b} \phi(r)\left[\frac{dk(r)}{dr}\right]\left[\frac{d\phi(r)}{dr}\right]^{-1} \tag{5}$$

$$k(r) = \sqrt{\frac{2\mu}{\hbar^2}\left[E_{nl} - V_{eff}(r)\right]} \tag{6}$$

where $r_a$ and $r_b$ are two turning points determined by $E = V_{eff}(r)$. $N = n+1$ is the number of the nodes of $\phi(r)$ in the region $E_{nl} = V_{eff}(r)$ and is larger by 1 than the $n$ of the nodes of the wave function $\psi(r)$. The first term $N\pi$ is the contribution from the nodes of the logarithmic derivatives of the wave function, and the second one is called the quantum correction. It was found that for all well-known exactly solvable quantum systems, this quantum correction is independent of the number of nodes of the wave function. This means that it is enough to consider the ground state in calculating the quantum correction $(Q_c)$, i.e.,

$$Q_c = \int_{r_a}^{r_b} k'_o(r)\frac{\phi_0}{\phi'_0}dr \tag{7}$$

To determine the energy eigenvalues we equate Eqs.(5) and (7).

### 3. Approximate solutions of the Schrödinger equation with extended Cornell potential

The SE for two particles interacting via potential $V(r)$ is given by [42].

$$\frac{d^2R(r)}{dr^2} + \frac{2\mu}{\hbar^2}\left[E_{nl} - V(r) - \frac{l(l+1)\hbar^2}{2\mu r^2}\right]R(r) = 0 \tag{8}$$

where $l, \mu, r$ and $\hbar$ are the angular momentum quantum number, the reduced mass for the quarkonium particle, inter-particle distance and reduced plank constant respectively.

We substitute Eq.(1) into Eq.(8) and obtain

$$\frac{d^2 R(r)}{dr^2} + \frac{2\mu}{\hbar^2}\left[E_{nl} - V_{eff}(r)\right]R(r) = 0 \tag{9}$$

where

$$V_{eff}(r) = \eta_0 r^2 + \eta_1 r - \frac{\eta_2}{r} + \frac{\eta_3}{r^2} + \frac{l(l+1)\hbar^2}{2\mu r^2} \tag{10}$$

We transform the coordinate of Eq.(9) by setting

$$x = \frac{1}{r} \tag{11}$$

Upon substituting Eq.(11) into Eq.(10) we have

$$V_{eff}(x) = \frac{\eta_0}{x^2} + \frac{\eta_1}{x} - \eta_2 x + \eta_3 x^2 + \frac{l(l+1)\hbar^2 x^2}{2\mu} \tag{12}$$

To deal with the first and second terms of Eq.(12), we propose the following approximation scheme.

We assume that there is a characteristic radius $r_0$ of the meson. Then the scheme is based on the expansion of $\frac{\eta_1}{x}$ and $\frac{\eta_0}{x^2}$ in a power series around $r_0$; i.e. around $\delta \equiv \frac{1}{r_0}$, up to the second order. This is similar to Pekeris approximation, which helps to deform the centrifugal term such that the modified potential can be solved by NU method [14].

Setting $y = x - \delta$ and around $y = 0$ it can be expanded into a series of powers as;

$$\frac{\eta_1}{x} = \frac{\eta_1}{y + \delta} = \frac{\eta_1}{\delta\left(1 + \frac{y}{\delta}\right)} = \frac{\eta_1}{\delta}\left(1 + \frac{y}{\delta}\right)^{-1} \tag{13}$$

which yields

$$\frac{\eta_1}{x} = \eta_1\left(\frac{3}{\delta} - \frac{3x}{\delta^2} + \frac{x^2}{\delta^3}\right) \tag{14}$$

Similarly,

$$\frac{\eta_0}{x^2} = \eta_0\left(\frac{6}{\delta^2} - \frac{8x}{\delta^3} + \frac{3x^2}{\delta^4}\right) \tag{15}$$

We then substitute Eqs.(14) and (15) into Eq.(12) and obtain

$$V_{\text{eff}}(x) = \xi_1 + \xi_2 x + \xi_3 x^2 \tag{16}$$

where

$$\left.\begin{aligned}\xi_1 &= \frac{6\eta_0}{\delta^2} + \frac{3\eta_1}{\delta}, \quad \xi_2 = \frac{3\eta_1}{\delta^2} - \frac{8}{\delta^3} - \eta_2 \\ \xi_3 &= \frac{l(l+1)\hbar^2}{2\mu} + \frac{3\eta_0}{\delta^4} + \frac{\eta_1}{\delta^3} + \eta_3\end{aligned}\right\} \tag{17}$$

The non-linear Riccati equation for ground state is written in terms of the new variable $x$ as,

$$-x^2 \phi'(x) + \phi^2(x) = k(x) \tag{18}$$

where

$$k(x) = \sqrt{\frac{2\mu}{\hbar^2}\left[\xi_3 x^2 + \xi_2 x + \xi_1 - E\right]} \tag{19}$$

We now apply the quantization rule to study the potential. To this end, we first calculate the turning points $x_a$ and $x_b$, which is determined by solving the content of the square bracket of Eq. (19), this yields;

$$\left.\begin{aligned}x_a &= \frac{-\xi_2 - \sqrt{\xi_2^2 - 4\xi_3(\xi_1 - E)}}{2\xi_3} \\ x_b &= \frac{-\xi_2 + \sqrt{\xi_2^2 - 4\xi_3(\xi_1 - E)}}{2\xi_3}\end{aligned}\right\} \tag{20}$$

From Eq. (20) we have

$$\left.\begin{aligned}x_a x_b &= \frac{\xi_1 - E}{\xi_3} \\ x_a + x_b &= -\frac{\xi_2}{\xi_3}\end{aligned}\right\} \tag{21}$$

Also, from Eq. (19) we have

$$k(x) = \sqrt{\frac{2\mu}{\hbar^2}\xi_3\left(x^2 + \frac{\xi_2}{\xi_3}x + \frac{\xi_1 - E}{\xi_3}\right)} \tag{22}$$

Substituting Eq.(21) into Eq.(22) we obtain

$$k(x) = \sqrt{\frac{2\mu\xi_3}{\hbar^2}(x - x_a)(x - x_b)} \tag{23}$$

where $k(x)$ is the momentum between the two turning points $x_a$ and $x_b$.

From Eq.(18), since the logarithmic derivative $\phi_0(x)$ for the ground state has one zero and no pole, therefore we assume the trial solution for the ground states

$$\phi_0(x) = A + Bx \tag{24}$$

Substituting Eq.(24) into Eq.(18) and then solving the non-linear Riccati equation, we obtain the ground state energy as

$$E_0 = \xi_1 - \frac{\hbar^2 A^2}{2\mu} \tag{25}$$

Also, we obtain $A$ and $B$ as follows

$$\left. \begin{array}{l} A = \dfrac{\mu \xi_2}{B\hbar^2} \\[2mm] B = \dfrac{1}{2} + \sqrt{\dfrac{1}{4} - \dfrac{2\mu \xi_3}{\hbar^2}} \end{array} \right\} \tag{26}$$

Here we choose the positive sign in front of the square root for $B$. This is as a result of the logarithmic derivatives $\phi_0(x)$ which decreases exponentially, which is required. We now calculate the quantum correction and obtain

$$\int_{r_a}^{r_b} \phi(r) \left[\frac{dk(r)}{dr}\right] \left[\frac{d\phi(r)}{dr}\right]^{-1} dr = -\int_{x_a}^{x_b} \frac{k_0'(x)}{x^2} \frac{\phi_0(x)}{\phi_0'(x)} dx \tag{27}$$

From Eq.(27) we have

$$= \sqrt{\frac{2\mu\xi_3}{\hbar^2}} \int_{x_a}^{x_b} \left( \frac{\dfrac{A}{B} - \dfrac{x_a + x_b}{2}}{x\sqrt{(x-x_a)(x-x_b)}} + \frac{\dfrac{A}{B}\left(\dfrac{x_a + x_b}{2}\right)}{x^2 \sqrt{(x-x_a)(x-x_b)}} \right) dx \tag{27a}$$

We utilized the integrals given by Appendix A, and obtain

$$= \pi\sqrt{\frac{2\mu\xi_3}{\hbar^2}}\left[\frac{\frac{A}{B}-\frac{x_a+x_b}{2}}{\sqrt{1+(x_a+x_b)+x_ax_b}}+\frac{\frac{A}{B}\left(\frac{x_a+x_b}{2}\right)\left(\sqrt{x_ax_b}-\frac{1}{2(x_a+x_b)}\right)}{\sqrt{x_ax_b}}\right] \qquad (27b)$$

We substitute Eq.(21) into Eq.(27b) and obtain

$$= \pi\sqrt{\frac{2\mu\xi_3}{\hbar^2}}\left[\frac{\frac{A}{B}+\frac{\xi_2}{2\xi_3}}{\sqrt{\frac{\hbar^2}{2\mu\xi_3}(A+B)}}-\frac{\frac{A\xi_2}{2B\xi_3}\left(\sqrt{\frac{\hbar^2}{2\mu\xi_3}}A+\frac{\xi_3}{2\xi_2}\right)}{\sqrt{\frac{\hbar^2}{2\mu\xi_3}}A}\right] \qquad (27c)$$

Inserting Eq.(27c) into Eq.(5) we obtain

$$= \pi\sqrt{\frac{2\mu\xi_3}{\hbar^2}}\left[\frac{\frac{A}{B}+\frac{\xi_2}{2\xi_3}}{\sqrt{\frac{\hbar^2}{2\mu\xi_3}(A+B)}}-\frac{\frac{A\xi_2}{2B\xi_3}\left(\sqrt{\frac{\hbar^2}{2\mu\xi_3}}A+\frac{\xi_3}{2\xi_2}\right)}{\sqrt{\frac{\hbar^2}{2\mu\xi_3}}A}\right]+N\pi \qquad (27d)$$

Furthermore, the integral of Eq.(7) is obtain as

$$\int_{r_a}^{r_b} k(r)dr = -\int_{x_a}^{x_b}\frac{k(x)}{x^2}dx \qquad (28)$$

From Eq.(28) we have

$$= -\sqrt{\frac{2\mu\xi_3}{\hbar^2}}\int_{x_a}^{x_b}\frac{\sqrt{(x-x_a)(x-x_b)}}{x^2}dx \qquad (28a)$$

Using Eq.(29), Eq.(28a) becomes

$$= -\sqrt{\frac{2\mu\xi_3}{\hbar^2}}\left[\frac{(x_a+x_b)-\sqrt{x_ax_b}}{2\sqrt{x_ax_b}}\right] \qquad (28b)$$

We substitute Eq.(21) into Eq.(28b) and obtain

$$-\sqrt{\frac{2\mu\xi_3}{\hbar^2}}\left[\frac{-\frac{\xi_2}{\xi_3}-\sqrt{\frac{\xi_1-E}{\xi_3}}}{2\sqrt{\frac{\xi_1-E}{\xi_3}}}\right] \tag{28c}$$

In order to obtain the integral of Eq.(28a), we used maple software to obtain the following useful integral, which is not available in integral table

$$\int_a^b \frac{\sqrt{(x-a)(b-x)}}{x^2}dx = \frac{(a+b)-\sqrt{ab}\pi}{2\sqrt{ab}} \tag{29}$$

By equating Eqs.(27d) and (28c), and substituting Eqs.(17) and (26), setting $\hbar=1$ we obtain the energy equation for extended Cornell potential as

$$E_{nl} = \frac{3\eta_1}{\delta}+\frac{6\eta_0}{\delta^2}-\frac{2\mu\left(\frac{3\eta_1}{\delta^2}+\frac{8\eta_0}{\delta^3}+\eta_2\right)^2}{\left[(2n+1)+\sqrt{1+\frac{8\mu\eta_1}{\delta^3}+4\left(\left(l+\frac{1}{2}\right)^2-\frac{1}{4}\right)-8\mu\eta_3+\frac{24\mu\eta_0}{\delta^4}}\right]^2} \tag{30}$$

3.1 Special case

In this subsection, we present some special cases of the energy eigenvalues of the ECP.

1. When we set $\eta_0 = \eta_1 = \eta_3 = 0$ we obtain energy eigenvalues expression for Coulomb potential

$$E_{nl} = -\frac{\mu\eta_2^2}{2(n+l+1)^2} \tag{31}$$

The result of Eq. (31) is consistent with the result obtained in Eq. (36) in Ref. [27]

2. When we set $\eta_0 = \eta_3 = 0$ we obtain energy eigenvalues expression for Cornell potential

$$E_{nl} = \frac{3\eta_1}{\delta}-\frac{2\mu\left(\frac{3\eta_1}{\delta^2}+\eta_2\right)^2}{\left[(2n+1)+\sqrt{1+\frac{8\mu\eta_1}{\delta^3}+4\left(\left(l+\frac{1}{2}\right)^2-\frac{1}{4}\right)}\right]^2} \tag{32}$$

The result of Eq. (32) is consistent with the result obtained in Eq. (30) in Ref. [14]

## 4. Results

Using the relation in Refs. [14, 18], we calculate the mass spectra of the heavy quarkonia such as charmonium and bottomonium, and heavy-light mesons.

$$M = 2m + E_{nl} \tag{33}$$

where $m$ is quarkonium bare mass and $E_{nl}$ is energy eigenvalues.

By substituting Eq.(30) into Eq.(33) we obtain the mass spectra for extended Cornell potential for heavy quarkonia as,

$$M = 2m + \frac{3\eta_1}{\delta} + \frac{6\eta_0}{\delta^2} - \frac{2\mu\left(\frac{3\eta_1}{\delta^2} + \frac{8\eta_0}{\delta^3} + \eta_2\right)^2}{\left[(2n+1) + \sqrt{1 + \frac{8\mu\eta_1}{\delta^3} + 4\left(\left(l+\frac{1}{2}\right)^2 - \frac{1}{4}\right) - 8\mu\eta_3 + \frac{24\mu\eta_0}{\delta^4}}\right]^2} \tag{34}$$

We also obtain the mass spectra of the special case of Eq.(32) for heavy-light mesons by substituting into Eq.(33)

$$M = m_s + m_c + \frac{3\eta_1}{\delta} - \frac{2\mu\left(\frac{3\eta_1}{\delta^2} + \eta_2\right)^2}{\left[(2n+1) + \sqrt{1 + \frac{8\mu\eta_1}{\delta^3} + 4\left(\left(l+\frac{1}{2}\right)^2 - \frac{1}{4}\right)}\right]^2} \tag{35}$$

Table 1

Mass spectra of charmonium in (GeV) ($m_c$ =1.209 GeV, $\mu$ = 0.6045 GeV, $\eta_1 = 0.001\ GeV$, $\eta_2 = 14.94\ GeV$, $\eta_0 = 0.02\ GeV =$, $\eta_3 = -15.04\ GeV$, $\delta = 1.7\ GeV$)

| State | Present work | AEIM[8] | NU[12] | AIM[19] | LTM[15] | SEM[35] | Experiment[43,44] |
|---|---|---|---|---|---|---|---|
| 1S | 3.096 | 3.0954 | 3.095 | 3.096 | 3.0963 | 3.095922 | 3.097 |
| 2S | 3.686 | 3.5673 | 3.685 | 3.686 | 3.5681 | 3.685893 | 3.686 |
| 1P | 3.295 | 3.5677 | 3.258 | 3.214 | 3.5687 | - | 3.525 |
| 2P | 3.802 | 4.0396 | 3.779 | 3.773 | 3.5687 | 3.756506 | 3.773 |
| 3S | 4.040 | 4.0392 | 4.040 | 4.275 | 4.0400 | 4.322881 | 4.040 |
| 4S | 4.269 | 4.5110 | 4.262 | 4.865 | 4.5119 | 4.989406 | 4.263 |
| 1D | 3.583 | 4.0396 | 3.510 | 3.412 | 4.0407 | - | 3.770 |

| State | Present work | | | | | | |
|---|---|---|---|---|---|---|---|
| 2D | 3.976 | - | - | - | - | - | 4.159 |
| 1F | 3.862 | - | - | - | - | - | - |

Table 2

Mass spectra of bottomonium in (GeV) ($m_b$ = 4.823 GeV, $\mu$ = 2.4115 GeV $\eta_1 = 0.798\ GeV$, $\eta_2 = 5.051\ GeV$, $\eta_0 = 0.02\ GeV =$, $\eta_3 = -3.854\ GeV$, $\delta = 1.5\ GeV$)

| State | Present work | AEIM[11] | NU[15] | AIM[18] | LTM[18] | SEM[35] | Experiment[43,44] |
|---|---|---|---|---|---|---|---|
| 1S | 9.460 | 9.74473 | 9.460 | 9.460 | 9.745 | 9.515194 | 9.460 |
| 2S | 10.023 | 10.02315 | 10.022 | 10.023 | 10.023 | 10.01801 | 10.023 |
| 1P | 9.661 | 10.02406 | 9.609 | 9.492 | 10.025 | - | 9.899 |
| 2P | 10.138 | 10.30248 | 10.109 | 10.038 | 10.303 | 10.09446 | 10.260 |
| 3S | 10.355 | 10.30158 | 10.360 | 10.585 | 10.302 | 10.44142 | 10.355 |
| 4S | 10.567 | 10.58000 | 10.580 | 11.148 | 10.580 | 10.85777 | 10.580 |
| 1D | 9.943 | 10.30248 | 9.846 | 9.551 | 10.303 | - | 10.164 |
| 2D | 10.306 | - | - | - | - | - | - |
| 1F | 10.209 | - | - | - | - | - | - |

Table 3

Mass spectra of bottom-charm ($b\bar{c}$) in (GeV) ($m_b$ = 4.823 GeV, $m_c = 1.209\ GeV$, $\eta_1 = 0.202\ GeV$, $\eta_2 = 1.213\ GeV$, $\delta = 0.371\ GeV$)

| State | Present work | AEIM[8] | NU[12] | LTM[15] | AIM[19] | Experiment[45] |
|---|---|---|---|---|---|---|
| 1S | 6.274 | 6.2774 | 6.277 | 6.2770 | 6.277 | 6.275 |
| 2S | 6.845 | 7.0376 | 7.383 | 7.0372 | 6.814 | 6.842 |
| 3S | 7.125 | 7.7978 | 7.206 | 7.7973 | 7.351 | - |
| 4S | 7.283 | 7.0386 | - | - | 7.889 | - |
| 1P | 6.519 | 7.7987 | 7.042 | 7.0381 | 6.340 | - |
| 2P | 6.959 | - | 6.663 | 7.7983 | 6.851 | - |
| 1D | 6.813 | - | - | - | 6.452 | - |

Table 4

Mass spectra of charm-strange (c$\bar{s}$) meson in (GeV) ($m_s = 0.419\ GeV$, $m_c = 1.209\ GeV$, $\eta_1 = 0.202\ GeV$, $\eta_2 = 2.046\ GeV$, $\delta = 0.561\ GeV$)

| State | Present work | NU[12] | AIM[19] | Experiment[43, 46] |
|---|---|---|---|---|
| 1S | 1.969 | 1.968 | 2.512 | 1.968[43] |
| 2S | 2.709 | 2.709 | 2.709 | 2.709[46] |
| 3S | 2.913 | 2.932 | 2.906 | - |
| 4S | 2.998 | - | 3.102 | - |
| 1P | 2.601 | 2.565 | 2.649 | - |
| 2P | 2.877 | - | 2.860 | - |
| 1D | 2.863 | 2.857 | 2.859 | 2.859[46] |

We test the accuracy of the predicted results determined numerically, by using the Chi square function defined by [37]

$$\chi^2 = \frac{1}{n} \sum_{i=1}^{n} \frac{\left(M_i^{Exp.} - M_i^{Theo.}\right)}{\Delta_i} \tag{36}$$

where $n$ runs over selected samples of mesons, $M_i^{Exp.}$ is the experimental mass of mesons, while $M_i^{Theo.}$ is the corresponding theoretical prediction. The $\Delta_i$ quantity is experimental uncertainty of the masses. Intuitively, $\Delta_i$ should be one. The tendency of overestimating Chi square value is that, it reflects some mean error per meson state.

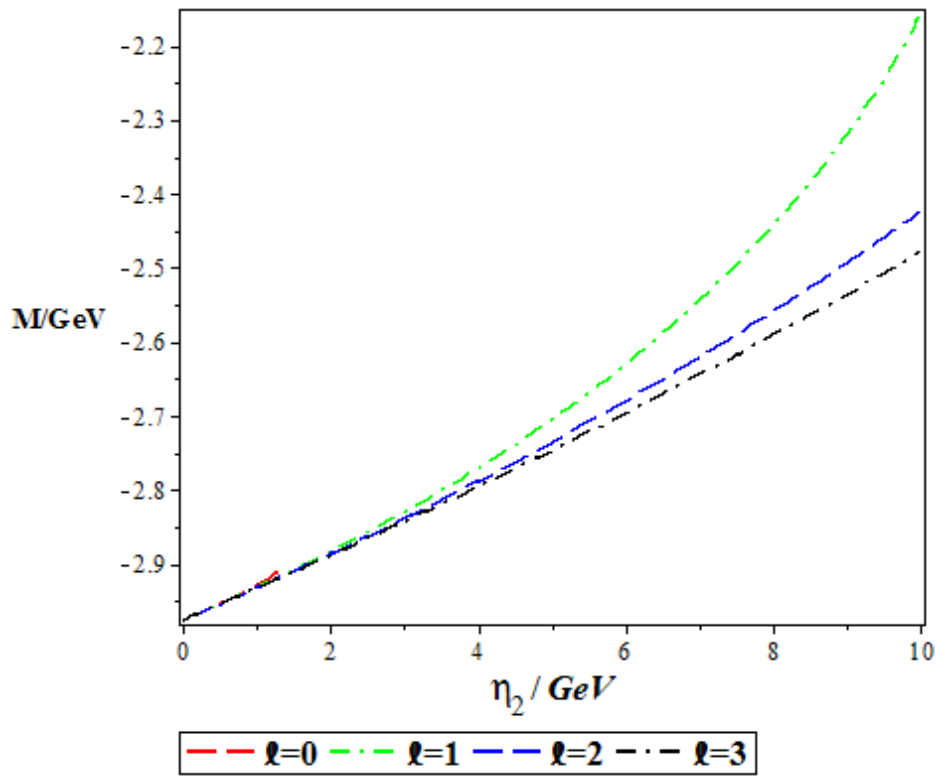

Figure 1: Variation of mass spectra with potential strength $(\eta_2)$ for different quantum numbers

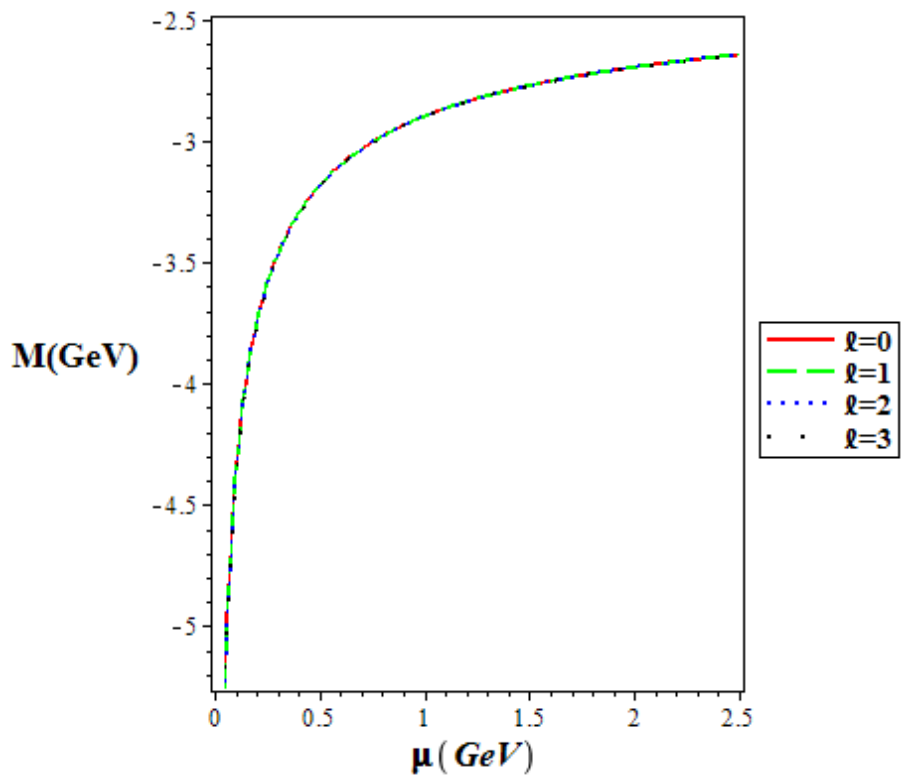

Figure 2: Variation of mass spectra with reduced mass $\mu$ for different quantum numbers

## 4.2 Discussion of results

We calculate mass spectra of charmonium and bottomonium for states from 1S to 1F, as presented in Tables 1 and 2. Also calculated the mass spectra of heavy-light mesons for states from 1S to 1D as presented in Tables 3 and 4. The free parameters of Eq.(34) were obtained by solving two algebraic equations of mass spectra for $2S, 2P$ in Eq.(34) for charmonium. We followed the same procedure for bottomonium and obtained the free parameters of mass spectra for $1S, 2S$ in Eq. (34). Equation (35) was fitted with experimental data of mass spectra of 1S, 2S to obtain the free parameters for bottom-charm $(b\bar{c})$ and charm-strange $(c\bar{s})$ heavy-light mesons.

For bottomonium $b\bar{b}$, charmonium $c\bar{c}$ and strange systems we adopt the numerical values of $m_b = 4.823\,GeV$, $m_c = 1.209\,GeV$ and $m_s = 0.419\,GeV$, and the corresponding reduced mass are $\mu_b = 2.4115\,GeV$, $\mu_c = 0.6045\,GeV$ and $\mu_s = 0.2095\,GeV$, respectively [43-46]. We note that the calculation of mass spectra of charmonium and bottomonium are in good agreement with experimental data and the work of other researchers like; Refs.[8,12,15,19,35] as shown in Tables 1 and 2. Also, the results for charm-strange meson presented in Table 4, are in excellent agreement with the work of other authors in Refs.[12,19] Furthermore, in Table 3, the mass spectra of the bottom-charm meson are very close to the ones obtained in Refs[8,12,15,19] with other methods and experimental data indicating an improvement compared to the other methods. The maximum error in comparison with experimental data is $0.0065\,GeV$. We also plotted mass spectra energy against potential strength and reduced mass, respectively. In Fig. 1, the mass spectra energy converges at the beginning but spread out and there is a monotonic increase in potential strength as the angular quantum number increases. Figure 2 shows the convergence of the mass spectra energy as the reduced mass $(\mu)$ increases for various angular quantum numbers.

## 5. Conclusion

In this study, we obtained the approximate solutions of the Schrödinger equation for energy eigenvalues with extended Cornell potential using the EQR approach. Two special cases were considered which result in Cornell and Coulomb potentials. We apply the present results to compute heavy-meson masses of charmonium and bottomonium and heavy-light mesons such as bottom-charm and charm-strange for different quantum states. We note that the mass spectra of the mesons systems obtained in this present work are improved and also in agreement with the results obtained by other researchers. This work could be extended to obtaining the thermodynamic properties of the mesons.

**Appendix A: Some Useful Standard Integrals**

$$\int_{r_a}^{r_b} \frac{1}{\sqrt{(r-r_a)(r_b-r)}} dr = \pi \tag{A1}$$

$$\int_{r_a}^{r_b} \frac{1}{(a+br)\sqrt{(r-r_a)(r_b-r)}} dr = \frac{\pi}{\sqrt{(a+br_b)(a+br_a)}} \tag{A2}$$

$$\int_{r_a}^{r_b} \frac{1}{r}\sqrt{(r-r_a)(r_b-r)} dr = \frac{\pi}{2}(r_a + r_b) - \pi\sqrt{r_a r_b} \tag{A3}$$

$$\int_{r_a}^{r_b} \frac{1}{r\sqrt{(r-r_a)(r_b-r)}} dr = \frac{\pi}{\sqrt{r_a r_b}} \tag{A4}$$

**Acknowledgement**

Dr. Etido P. Inyang would like to thank C. O. Edet, Department of Physics, University of Port Harcourt, Nigeria for his encouragement for the successful completion of this work.

**References**

[1] R. Kumar, F. Chand, Commun. Theore. Phys. 59, 456 (2013)

[2] H. Mutuk, Advan. in High Energy Phys.20, 653 (2018)

[3] M.N.Sergeenko, (2019),https://arxiv.org/abs/1909.10511

[4] C. Cari, A. Suparmi, M.Yunianto, B.N.Pratiwi, J. Phys. Con. Series 776, 92 (2016)

[5] M. Abu-shady,Int. J. Appl. Math. and Theor. Phys. 2(2), 16 (2016)

[6] H. Ciftci, H.F. Kisoglu, Advan. High Energy Phys.45, 497 (2018)

[7]A. Vega and J.Flores, Pramana J.Phys. 87 (2016)

[8] E.M.Khokha,M. Abushady, T.A. Abdel-Karim, Int. J. Theore. Appl. Math. 2, 86 (2016)

[9] A. Al-Jamel, H. Widyan, Appl. Phys. Research 4 (2012)

[10] E.Omugbe, Cana. J. Phys. 45 (2020)

[11]H. Mutuk, Advan. High Energy Phys. (2019) 31, 373 (2019)

[12] M. Abu-Shady,T.A. Abdel-Karim, Y. Ezz-Alarab, J. Egypt. Math. Soc. 27, 155 (2019)

[13] H. Mansour, A. Gamal, Advan. High energy. Phys.65, 1234 (2018)

[14] M. Abu-Shady, J. Egypt. Math. Soc. 23, 4 (2016)

[15] M. Abu-Shady, T.A. Abdel-Karim, E.M. Khokha, J. of Quantum Phys. 45, 567 (2018)

[16] M. Abu-Shady, Boson J. Mod. Phys. 1(1), 16 (2015)

[17] S.M.Ikhdair,Advan. High Energy Phys. 49, 648 (2013)

[18] E.P. Inyang , E.P. Inyang, J.E. Ntibi, E.E. Ibekwe, E.S. William, Indian J. Phys. 20, 00097R2(2020)

[19] R. Rani, S. B. Bhardwaj, F.Chand, Commun.Theore. Phys. 70, 179(2018)

[20] R.L.Hall N.Saad, Open Phys. 13, 23 (2015)

[21] M. Abu-Shady, A.N. Ikot, Euro. Phys. J. 54, 134 (2019)

[22] A, Al-Jamel, Int. J. Mod. Physics, 34, 242 (2019)

[23] C.O.Edet, P.O. Okoi, Revi. Mexi. Fisica 65, 333(2019)

[24] J. E. Ntibi, E.P.Inyang, E.P.Inyang, E.S.William, Int. J. Innovative Science, Engineering & Tech. 11 2348(2020)

[25] J.A.Obu, E.S.William,I.O.Akpan,E.A.Thompson,E.P.Inyang,Euro.J.Appl.Phys.2(2020)

[26] C.O.Edet, U.S.Okorie, G. Osobonye, A.N. Ikot, G.J. Rampho, R.Sever , J. Math. Chem.18 (2020)

[27] C.O.Edet, U.S. Okorie, A.T.Ngiangia, A.N.Ikot, Indian J. Phys. 94, 423 (2020)

[28] P.O.Okoi, C.O.Edet, T.O. Magu, Rev. Mexi. de Fis. 66, 6(2020)

[29] C.O.Edet, P.O. Okoi, A.S.Yusuf, P.O.Ushie, P.O.Amadi, Indian J. phys. 8 (2020)

[30] A.N.Ikot,C.O.Edet,P.O.Amadi,U.S.Okorie, G.J.Rampho, H.Y.Abdullah, Euro. Physi. J. D, 74 (2020)

[31] E.P.Inyang, J. E.Ntibi, E.P.Inyang, E.S.William, C.C. Ekechukwu, Int. J. Innovative Science, Engineering &Tech. 11, 2432(2020)

[32] C.O.Edet, P.O. Okoi, S.O.Chima, Rev. Brasil. Ensi. Fisica. 8 (2019)

[33] C.O.Edet, P.O. Amadi, M.C.Onyeaju, U.S.Okorie,R. Sever, G.J.Rampho, J. Low Temp. Phys. 9 (2020)

[34] E.P.Inyang,E.P.Inyang,I.O.Akpan, J.E.Ntibi,E.S.William,Euro.J.Appl.Phys. 2 26(2020)

[35] E.E .Ibekwe , T.N.Alalibo, S.O.Uduakobong, A.N.Ikot, N.Y.Abdullah, Iran J. Sci. Tech. 20, 913 (2020)

[36] E. Omuge, O.E.Osafile, M.C.Onyeajh, Adv. High Ener. Phys.10, 1143 (2020)

[37] M.S.Ali, A.M.Yasser,G.S.Hassan, C.C.Moustakidis,Quan.Phys. Lett.5 ,14 (2016)

[38] A.Vega, J. Flores, Pramana-J.Phys.56, 456 (2016)

[39] A. Al-Oun, A. Al-Jamel, H. Widyan, Jordan J. Phys.40, 453 (2015)

[40] Z.Q. Ma, B.W.Xu, Europhys.Lett.69, 685 (2005)

[41] Z.Q. Ma, and B.W.Xu, Int.J.Mod.Phys.E 14, 599 (2005)

[42] E.S.William ,E.P. Inyang, E.A.Thompson, Rev. Mexi. Fisica. 66 730 (2020)

[43] R.Olive, D.E.Groom, T.G.Trippe, Chin. Phys. C,38 ,54 (2014)

[44]M. Tanabashi, C,D,Carone, T.G.Trippe, C.G.Wohl, Phys. Rev. D, 98, 546 (2018)

[45] C. Patrignani, Chin. Phys. C 40 ,100 (2016)

[46] S. Godfrey, K. Moats, (2015) https://arxiv.org/abs/1409.0874v3.